\documentclass[twocolumn]{aa} 
\usepackage{amsmath}
\usepackage{amsfonts}
\usepackage{amssymb}
\usepackage{graphicx,xcolor}
\usepackage{algorithm,algorithmic}
\usepackage{txfonts}
\begin{document}

\title{Fast simulations of extragalactic microlensing\thanks{Find simulator and code at https://microlensing.overfitting.es}}

\titlerunning{Fast microlensing simulations}

\author{V. N. Shalyapin\inst{1,2,3} 
\and
R. Gil-Merino\inst{1,4}
\and
L. J. Goicoechea\inst{1}}

\institute{GLENDAMA Team, Dpto. de F\'\i sica Moderna, Universidad de Cantabria, Avda. de Los Castros s/n, E-39005 Santander, Spain\\
\email{vshal@ukr.net;gilmerino@uma.es;goicol@unican.es}
        \and
O.Ya. Usikov Institute for Radiophysics and Electronics, National Academy of Sciences of Ukraine, 12 Acad. Proscury St., UA-61085 Kharkiv, Ukraine
        \and
Institute of Astronomy of V.N. Karazin Kharkiv National University, Svobody Sq. 4, UA-61022 Kharkiv, Ukraine
        \and
Dpto. de Lenguajes y Ciencias de la Computaci\'on, E.T.S.I Inform\'atica, Campus de Teatinos, Universidad de M\'alaga, Bulevar Louis Pasteur 35, E-29071 M\'alaga, Spain
}


\abstract{We present a new and very fast method for producing microlensing magnification maps at high optical depths. It is based on the combination of two approaches: (a) the two-dimensional Poisson solver for a deflection potential and (b) inverse polygon mapping. With our method we extremely reduce the computing time for the generation of magnification patterns and avoid the use of highly demanding computer resources. For example, the generation of a magnification map of size $2000 \times 2000$ pixels, covering a region of 20 Einstein radii, takes a few seconds on a state-of-the-art laptop. The method presented here will facilitate the massive production of magnification maps for extragalactic microlensing studies within the forthcoming surveys without the need for large computer clusters. The modest demand of computer power and a fast execution time allow the code developed here to be placed on a standard server and thus provide the public online access through a web-based interface.}


\maketitle

\section{Introduction}\label{sec:intro}
Gravitational lensing occurs when a mass distribution lenses a background source (\citealp{1992grle.book.....S} and references therein). Substructure in the lens mass distribution, in granular or/and smooth form, might induce gravitational microlensing in the received images of the farther source \citep{1964PhRv..133..835L, 1979Natur..282..561, 1986ApJ...301..503P}. Typically, at cosmological distances and in a strong lensing regime, the background source is a distant galaxy or a quasar. The lens is a much closer galaxy that bends the light, producing multiple images of the source. The substructure that induces microlensing is formed by the stars and the interstellar medium within the lens galaxy \citep{1989AJ.....98.1989I, 1991AJ....102...34C, 1994ApJ...429...66W}. Substructure can be detected by comparing the different light curves of the multiple lensed images. The study of both lensed and microlensed systems is a tool for gaining insights into the nature, and unresolved structure, of the background sources as well as into the mass distribution of the lensing galaxies \citep{1990ApJ...352..407W, 1991AJ....102.1939W, 1992ApJ...386L..43F, 1998A&A...335..379S, 1998Ap&SS.263...47G, 1999ApJ...519L..31Y}.

The study of gravitational microlensing is done by using magnification patterns, that is, the projection of the light that crosses the lens plane onto the source plane. The variables used to define the properties of the lens plane are: the distribution of matter in compact objects, the smoothly distributed matter, and the gravitational shear. The influence of the source in the magnification pattern is introduced by a convolution with the corresponding source profile. Straight lines in the resulting magnification patterns mimic the movement of the source for a given period of time, producing synthetic light curves that can be directly compared to the observed light curves. Microlensing-induced features in observed spectra can also be interpreted in terms of concentric sources with different shapes and sizes around points in magnification patterns. The random motion of the stars in the lens plane can additionally be taken into account, increasing the complexity of the simulations and the number of magnification patterns needed for a given configuration. The statistical properties of a number of synthetic light curves put limits on the parameters used in the simulations in order to be consistent with the observed data.

Several methods have been proposed to produce such magnification patterns in the strong lensing regime, the most popular being the inverse ray-shooting (IRS) method (\citealp{1986A&A...166...36K, 1986ApJ...301..503P, 1987A&A...171...49S, 1990PhDT.......180W}). The IRS method consists in shooting rays backwards from a regular grid of rays at the lens plane to the source plane. The magnification in a given pixel of the source plane is then proportional to the number of rays that hit it. The number of calculations in the IRS method is proportional to both the number of rays shot and the number of stars on the lens plane. The IRS method uses, as standard numbers, $10^3-10^4$ stars and $\geq 100$ rays per pixel, which implies demanding computational requirements when the number and dimensions of the magnification patterns are large.

Mathematically, the parameters to produce a magnification pattern and the lens equation are related in the following way: If the convergence due to compact objects is $\kappa _{\star}$, the convergence due to smoothly distributed matter is $\kappa_\textnormal{s}$, the gravitational shear is $\gamma$, and the scaled deflection angle is $\boldsymbol{\alpha(\boldsymbol{x})}$, then a point at position $\boldsymbol{x}$ in the lens plane is related to the point at position $\boldsymbol{y}$ in the source plane through the lens equation
\begin{equation}\label{eq:lenseq}
\boldsymbol{y} = 
\begin{pmatrix} 
1-\kappa _\textnormal{s}-\gamma & 0\\ 
0 & 1-\kappa _\textnormal{s}+\gamma 
\end{pmatrix}
\boldsymbol{x} - \boldsymbol{\alpha(\boldsymbol{x}})\, ,
\end{equation}
and the scaled deflection angle can be written as
\begin{equation}\label{eq:angle}
\boldsymbol{\alpha(\boldsymbol{x}}) = \sum_{l=1}^{N_\star}m_l \frac{\left ( \boldsymbol{x - x_l} \right )}{\left | \boldsymbol{x -\boldsymbol{x_l}} \right | ^{2}},
\end{equation}
where $m_l=M_l/\langle M\rangle$ is the relative mass and $\boldsymbol{x_l}$ the position of each microlens. In Eqs.~(\ref{eq:lenseq}) and (\ref{eq:angle}), $\boldsymbol{x}$ and $\boldsymbol{y}$ are dimensionless vectors since they were scaled by taking a lens-plane length equal to the Einstein radius for a point-like object with the mean microlens mass $\langle M\rangle$ and the corresponding source-plane length (e.g. Schneider et al. 1992). From here on, we generally refer to microlenses as stars, regardless of whether or not they really are stars.

The number of calculations needed to solve these equations at high optical depth in the lens plane ($N_\textnormal{calc}$), that is, for a large number of microlenses, is proportional to the number of stars ($N_\star$), the number of cells ($N_\textnormal{cell}$), and the number of rays inside of a single cell ($N_\textnormal{ray}$) in the following manner:
\begin{equation}\label{eq:dependences}
N_\textrm{calc} \sim N_\textrm{cell} \cdot N_\star \cdot N_\textrm{ray}.
\end{equation}
The problem gains in complexity when analysing light propagation through many deflection planes or when taking random velocities of stars into account.

In recent years, most efforts to produce magnification patterns in a quicker manner have focused on the available computing power. The work by \cite{2010NewAstr..15..725} compares a ray-shooting implementation on a graphical processing unit (GPU) to a hierarchical tree code on a single-core central processing unit (CPU) and a parallel tree code running on a cluster of CPUs. The project GERLUMPH \citep{2013MNRAS.434..832V} is a paradigmatic case, with a large number of GPUs used to reduce processing time. In the search for an alternative to the GPUs, \cite{2017Astr.Comp..19..60} sought a way to accelerate the microlensing simulations with the use of the Xenon Phi co-processor. Other, less explored ways of reducing computing time in the generation of magnification patterns are the search for new and faster mathematical approaches, either reducing the dependence on the number of stars, $N_\star$, or on the number of rays, $N_\textnormal{ray}$, needed in the simulations.

The reduction in the dependence on the number of stars can be done in several ways. For example,  \cite{1987A&A...171...49S} expanded the deflection angle of all distant stars up to the second order. \cite{1990PhDT.......180W} combined the microlenses in groups with larger sizes for more distant stars in a hierarchical tree code. Finally, \cite{1995ARep...39..594S} proposed the solution of the two-dimensional Poisson equation and explored models with high star density \citep{2006AN....327..981P}.

Reducing the dependence on the number of rays was successfully explored by \cite{2006ApJ...653..942M}. These authors suggested using unit cells, instead of separate rays, on the lens plane, to be transformed by gravitational deflection. The  magnification is then defined by the ratio of the transformed cell area to the original unit cell area, instead of using the count number of rays in the source plane. This method, named inverse polygon mapping (IPM), allows the number of rays per pixel to be drastically reduced, from several hundred to only a few. In fact, even one deflection angle calculation per pixel in IPM provides an accuracy comparable to that of several hundred calculations of deflection angles in the IRS method (see Sect. 6 in \citealp{2006ApJ...653..942M}).

To our knowledge, the attempts to reduce the dependences on $N_\star$ and $N_\textnormal{ray}$ have been done separately in different methods. To date, no method has attempted to combine a reduction in the dependences on $N_\star$ and $N_\textnormal{ray}$ simultaneously. In the present work, we present such a combined approach: Sect.~\ref{sec:poisson} briefly outlines the method of reducing the dependence on $N_\star$ that was already published in \cite{1995ARep...39..594S}. Section~\ref{sec:ipm} describes the IPM method from \cite{2006ApJ...653..942M}, which practically eliminates the dependence on $N_\textnormal{ray}$. Section~\ref{sec:combined} presents the combination of the two aforementioned approaches and analyses the accuracy and efficiency of our new combined method compared to the IPM method, on which our method is based. Section~\ref{sec:web} is devoted to the web-based tool: The new combined method is already implemented in a publicly accessible online simulator\footnote{\url{https://microlensing.overfitting.es}}, where the user can generate a magnification map of a maximum size of $2000 \times 2000$ pixels, covering a maximum area of $100$ Einstein radius, and then download it as a binary file for further analysis (other sizes are easily available on demand). Finally, Sect.~\ref{sec:final} presents our conclusions, a discussion, and future perspectives.

\section{The two-dimensional Poisson equation for a deflection potential}\label{sec:poisson}
The deflection angle in Eq.~(\ref{eq:angle}) can be directly calculated as a summation of the gravitational contribution of every microlens. Alternatively, the deflection angle, $\boldsymbol{\alpha}$, can also be computed from the gradient of the two-dimensional deflection potential, $\psi(\boldsymbol{x}),$ as
\begin{equation}\label{eq:gradient}
\boldsymbol{\alpha}(\boldsymbol{x}) = \triangledown \psi(\boldsymbol{x}).
\end{equation}
The gradient here is contained in the image plane. By taking the integral of the deflection angle in Eq.~(\ref{eq:angle}), we obtain an expression for the deflection potential, $\psi(\boldsymbol{x}),$ as
\begin{equation}\label{eq:potential}
\psi(\boldsymbol{x}) = \sum_{l=1}^{N_\star}m_l \ln \left | \boldsymbol{x} -\boldsymbol{x_l} \right |.
\end{equation}

The two-dimensional Poisson equation relates this deflection potential and the convergence due to microlenses with the equation:
\begin{equation}\label{eq:poisson_k}
\triangledown^2 \psi(\boldsymbol{x}) = 2 \kappa_\star(\boldsymbol{x})
,\end{equation}
where $\kappa_\star = \sum_{l=1,N_\star} \pi~m_l~\delta^2(\boldsymbol{x}-\boldsymbol{x_l})$ and $\delta^2$ is the two-dimensional Dirac delta function  \citep*{1992grle.book.....S}. To solve the two-dimensional Poisson equation (Eq. \ref{eq:poisson_k}), we considered a discrete mesh formed by square grid cells of side $h$ along $x_1$ (horizontal) and x$_2$ (vertical) directions. For the grid node at x = ($x_{1i}$, $x_{2j}$), a finite difference technique allowed us to replace $\triangledown^2 \psi$ with an equivalent finite difference quotient. Additionally, as the convergence $\kappa_\star$ is not a smooth function, we assigned an effective mass to the grid node, weighting the contribution of point-like masses within the region R around it. At this node $(i, j)$, we obtain the deflection potential-effective mass relationship:
\begin{equation}\label{eq:Poisson}
\psi(i+1,j)+\psi(i-1,j)+\psi(i,j+1)+ \psi(i,j-1)-4\psi(i,j) = 2 \pi m(i,j),\footnotemark
\end{equation}
\footnotetext{This expression can be written by dividing both sides of the equation by the factor $h^2$. The physical interpretation is then easier: The left side is the Laplacian of the potential, and the right side is twice the convergence due to stars. We, however, keep the more compact form in the main text.}where the effective mass assigned to the node, $m(i,j) = \sum_{l \in R} m_l W[(x_l-x)/h]$, is due to all the microlenses within the region R, and the weight function W depends on the node-microlens scaled separation (e.g. \citealp{1988csup.book.....H, 2003astro.ph..4162B}).

To estimate the effective masses at grid nodes, we used the `cloud-in-cell' method \citep*{1988csup.book.....H}. This method splits the mass of a microlens inside a grid cell into four smaller masses at the four grid nodes of the cell, according to the separations between the microlens and the nodes. Thus, the mass $m(i,j)$ comes from the  microlenses inside the four cells closest to the node $(i, j)$. Although we adopted the cloud-in-cell method for assigning masses to grid nodes, other schemes are also possible. For instance, the `triangular shape cloud' technique distributes the mass in a smoother way. Unfortunately, this method leads to relatively smooth variations in the microlensing magnification, which cannot account for certain high magnification events.

In order to solve the Poisson equation, boundary conditions at the edges of the shooting rectangular regions are also needed. There are two standard ways to include these boundary conditions: (a) Dirichlet conditions on the potential values in Eq.~(\ref{eq:Poisson}); and (b) Neumann conditions on the derivatives of the potential, that is, the scaled deflection angle in Eq.~(\ref{eq:angle}). Although both approaches work well, when using Neumann boundary conditions the potential is defined, except for an arbitrary constant that has no impact in determining the deflection angle. To obtain boundary conditions, we used Eq.~(\ref{eq:potential}) (the Dirichlet conditions) at the boundary points. The computing time required for these calculations depends on the number of points at the boundaries, but there are always fewer points than those in an entire calculation region (more details on the computing time are given in Sect.~\ref{sec:combined}). Equation~(\ref{eq:Poisson}) with boundary conditions can be solved in an effective way using the fast Fourier transformation methods described in \cite{1995ARep...39..594S} or using an optimised numeric library, for example the Fast Poisson Solver from the Intel Math Kernel Library.

The components of the scaled deflection angle were calculated by replacing Eq.~(\ref{eq:gradient}) with two equivalent finite difference equations, namely 
\begin{equation}
\alpha_{x1}(i,j) = \left[\psi(i+1,j)-\psi(i-1,j)\right]/(2h)
\end{equation}
and
\begin{equation}
\alpha_{x2}(i,j) =  \left[\psi(i,j+1)-\psi(i,j-1)\right]/(2h).
\end{equation}
From the potential values, we also computed the second partial derivatives of the deflection potential at mesh points and subsequently estimated  the magnification inside cells centred on these points.

\section{Inverse polygon mapping}\label{sec:ipm}
The IPM method considers a periodic lattice of rays that tessellates the entire image plane. The lattice can be seen as a large set of vertices of congruent polygons that are called unit cells. These unit cells are mapped onto the source plane by the lens equation. The lens mapping is an invertible linear transformation (except when cells are crossed by critical curves), so polygons in the image plane are transformed (mapped) into polygons in the source plane, which are mostly highly distorted. The IRS method relies on collecting light rays in pixels (unit cells) in the source plane, while the IPM technique consists in collecting image-plane unit cells into the source-plane pixels.

Although \cite{2006ApJ...653..942M} described the IPM technique, their algorithm for apportioning the mapped polygons among source-plane pixels is not public. Therefore, we developed our own methodology to calculate the intersecting area of a mapped polygon (parallelogram) and a source-plane pixel (square) partially covered by it. We closely followed the ideas in \cite{1992ACM...11.3..276} and found that our procedure is significantly faster than that proposed by \cite{2006ApJ...653..942M} because ours does not depend on the number of stars, $N_\star$ (see Table 1). In Appendix~\ref{ap:intersect} we illustrate in detail how our method works.

We determined the points at which the sides of both geometrical objects, the transformed polygon and the source-plane unit cell, intersect. These points, together with the unit cell vertices covered by the polygon and the polygon vertices inside the source-plane pixel, define the set of reference points. In a final step, we used the full set of $N$ reference points ($N$ $\le$ 8), sorted in anticlockwise order, to build an intersection polygon with $N$ vertices at $\boldsymbol{y}^i=(y_1^i, y_2^i)$, $i = 1, …, N$. In the example that we show in Fig.~\ref{fig:intersect} of Appendix~\ref{ap:intersect}, the intersection polygon has five vertices ($N$ = 5; see the small green squares). The area of any convex intersection polygon with N anticlockwise-ordered vertices is given by
\begin{equation}\label{eq:convexarea}
S = \frac{1}{2} \sum\limits_{i=1}^{N-1}\left(y_1^i y_2^{i+1} - y_1^{i+1}y_2^i\right) + \frac{1}{2} \left(y_1^N y_2^1 - y_1^1 y_2^N\right).
\end{equation}

In this paper we used a simple variant of the IPM technique, without a control to detect out-of-linearity cells in the image plane. Although it is possible to improve this simple scheme by identifying out-of-linearity cells (when they are crossed by critical curves) and breaking them into smaller cells \citep{2006ApJ...653..942M, 2011ApJ...741...42M}, we did not consider these sophisticated variants because we achieved a very good compromise between accuracy and execution time (see Sect.~\ref{sec:combined}).

\section{The combined method: Poisson solver and inverse polygon}\label{sec:combined}
In this section we compare the accuracy and computing time requirements of the combined method presented in this work, that is, the combination of the two-dimensional Poisson solver and the IPM method. We call the combined method the PIP (Poisson and inverse polygon) method. Our method provides an improvement in computing time over the IPM method without a loss of accuracy. For this reason, we compare our method directly to the IPM method here, and we and refer the reader to \cite{2006ApJ...653..942M} to consider the comparison to other methods where IPM has already shown advantages.

The IPM method produces excellent accuracy even in the simple variant without control of critical cells and with only one ray per un-lensed pixel. Its computing time compared to the IRS method is also very significantly reduced (see \citealp{2006ApJ...653..942M} for a detailed time comparison).

\begin{figure*}
\begin{center}
\includegraphics[width=\textwidth]{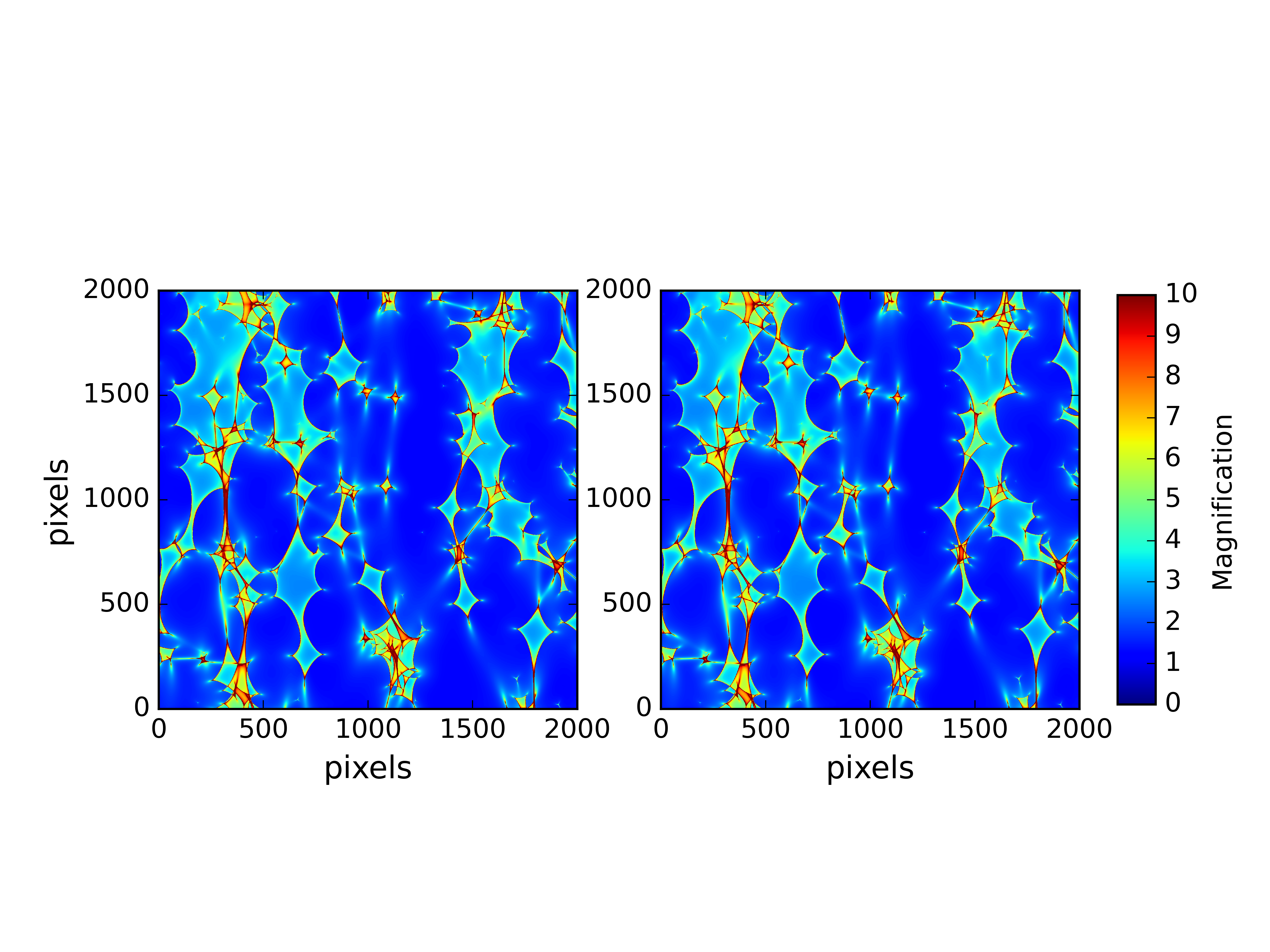}
\end{center}
\caption{Magnification maps produced for $\kappa_{\star},\gamma, \kappa_{\textnormal{s}} = (0.3,0.3,0.0)$, with a width of $20 R_{\rm E}$ and a resolution of $2000 \times 2000$ pixels. The left panel was produced with the IPM method and the right one with our PIP method.}
\label{fig:maps}
\end{figure*}

For a test case, we chose a typical configuration with convergence ($\kappa_\star$) and shear ($\gamma$) equal to 0.3 and all mass in stars (i.e. $\kappa_{\textnormal{s}} = 0$). Using a magnification map of size $2000 \times 2000$ pixels, a shooting region in the image plane occupies $7500 \times 3500$ cells, that is,  1.5 $\times$ $N_\textnormal{pix}$/(1$-\kappa-\gamma$) = 7500 and 1.5$ \times$ $N_\textnormal{pix}$/(1$-\kappa+\gamma$)= 3500. This approach was also used in the IPM code by Mediavilla and collaborators. In Fig.~\ref{fig:maps}, with $N_\star$ = 537 (see the third row in Table~\ref{table1}), the left panel shows a map obtained using the IPM method and the right using our PIP method. The standard deviation of the relative difference between the two maps (i.e. $\sigma$[(PIP-IPM)/IPM]) is 1.2\%. This deviation is defined by only a few pixels because mean and median values of relative difference modulo are 0.29\% and 0.13\%, respectively.

\begin{figure*}
\sidecaption
 \includegraphics[width=12cm]{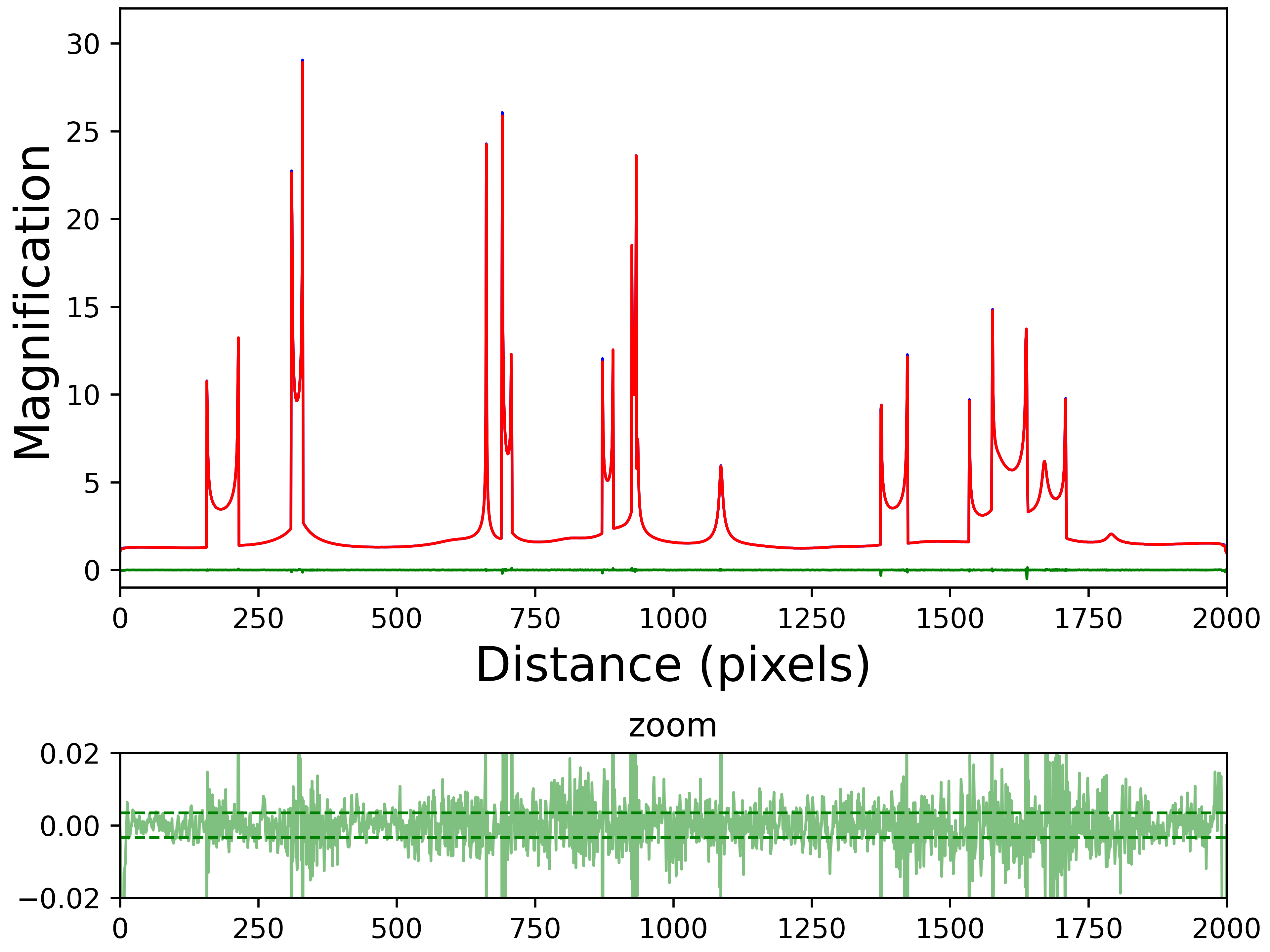}
  \caption{Light curves along the middle row of the maps shown in Fig.~\ref{fig:maps} (horizontal path at pixel 1000 on the vertical axis). Top panel: Light curve for IPM drawn in blue, with the light curve of the combined method superimposed and plotted in red. Both curves are practically indistinguishable, and their differences (in green) are close to zero. Bottom panel: Relative differences between the two light curves. The differences are normalised with the IPM map values. The relative difference signal has a noisy behaviour around zero, and the mean value of their absolute values is only 0.0029, i.e. 0.29\%.}
  \label{fig:tracks}
\end{figure*}

We can also compare the tracks along the magnifications patterns, which can be considered as the flux variations of the source in time. A comparison of tracks for a randomly selected path across the magnification maps is shown in Fig.~\ref{fig:tracks}. There is a minimal difference at the top of the caustic crossing peaks that can be neglected without losing any features in the light curves. These little differences come from the use of a discrete Poisson equation that relies on the assignation of effective masses to grid nodes through the cloud-in-cell technique. The two-dimensional Poisson solver is accurate, but the effective mass distribution is different from (although based on and consistent with) the simulated distribution of masses. The difference is $1\%$ when considering the whole magnification map. Whether or not this difference is acceptable depends on the particular application, but the difference can be reduced by decreasing the cell size.

As for the computing time, we kept the magnification map size as $2000 \times 2000$ pixels and the microlensing parameters set ($\kappa_\star,\gamma, \kappa_{\textnormal{s}}$) as in the previous case, but we checked different sizes for the physical magnification maps in Einstein radii. The size of the shooting region is still $7500 \times 3500$ image-plane unit cells (pixels), but the number of stars varies accordingly. All results were obtained with an Intel i5-2300 CPU @ 2.80 GHz with four cores (Table \ref{table1}).
\begin{table}
\caption{Computing time (in seconds) for different source regions (with length L in $R_E$) and numbers of stars, $N_\star$.}
\label{table1}
\centering
    \begin{tabular}{c c c c}
    \hline\hline            
            L & $N_\star$ & $t_\textrm{PIP}(t_\textrm{pot})$ & $t_\textrm{IPM}$\\
    \hline
            5        & 33       & 3  (0.35)  & 17     \\
            10       & 134      & 7 (0.35)  & 56     \\
            20       & 537      & 13 (0.41)  & 191    \\
            40       & 2148     & 28 (0.54)  & 563    \\
            80       & 8594     & 45 (1.04)  & 1722   \\
            160      & 34377    & 34 (3.03)  & 6126   \\
            320      & 137509   & 30 (11.0)  & 23550  \\
         \hline
     \end{tabular}
     \tablefoot{The third column indicates the time required by our new combined approach, $t_\textnormal{PIP}$, the fourth column shows the time of the combined method spent for the  potential calculation, $t_\textnormal{pot}$, and the last column provides the time required by the IPM method, $t_\textnormal{IPM}$.}
   \end{table}

Our PIP method calculates the magnification maps in two steps: First, it calculates the potential and, second, the intersections of the transformed unit cells with the source unit cells. This is the reason we include the time spent only for the potential calculation in the fourth column of Table~\ref{table1} (within parentheses).  The last column provides the time required by the IPM method. Comparison of the computing times of our PIP method and the IPM method shows a moderate advantage of the former over the latter for small numbers of stars, but the advantage becomes significantly bigger at large optical depths. In addition, the computing time of our PIP method slightly increases when the size of the source region increases; however, after reaching a maximal value, it does not increase for a higher number of stars.

This last effect can be explained by the joint properties of our combined algorithm. The time spent for the potential calculation is increased for large numbers of stars  because one needs to add their contribution in the direct boundary condition calculations. At the same time, for a given physical size in pixels of the magnification map, if we increase the area of the magnification map in Einstein radii, then one Einstein radius covers fewer cells. This means that the algorithm spends less time in apportioning the mapped polygons among a lower number of source plane pixels. The same effect is inherent for the IPM method as well, but in that case the deflection angles are calculated at each source cell, rather than only at the boundaries, and the time required to compute the deflection angle increases.

The PIP method only spends a small amount of time on calculating the boundary conditions. As an example, in the third row of Table~\ref{table1}, $t_\textnormal{PIP}$ = 13 s, $t_\textnormal{pot}$ = 0.41 s, and the boundary conditions evaluation takes only $\sim$0.1 s. Therefore, calculations of boundary conditions for finite star fields are fast enough to not consider other alternatives. Instead of finite star distributions with boundary conditions, we could have used periodic star fields (e.g. \citealp{2004ApJ...605...58K}). However, this alternative is not suitable for analysing the microlensing signal produced by a specific distribution of stars in a given region.

\section{Algorithm and implementation of the web-based solution}\label{sec:web}
The importance of magnification maps in quasar microlensing research has already been briefly reviewed in Sect.~\ref{sec:intro} (see references therein). The evolution of the methods in the generation of magnification maps has been closely associated with the evolution of supercomputers. In the present paper we develop a new method that changes the paradigm of microlensing magnification pattern calculation: It does not require sophisticated computer resources, and the magnification maps can be generated using a standard home computer in only a few seconds.

The modest requirements of our PIP method presented here, together with the extremely short execution time, allow a production environment to be allocated at a remote server with a client-server architecture. In this way, the user does not need to install or run any additional software at his/her local computer. In fact, the user does not even need a computer; any gadget with internet access would be sufficient to explore microlensing maps. This feature could significantly extend the number of potential users. Due to the very fast calculation method, the user interacts with the server calculations in real time.

The extragalactic microlensing magnification map generator installed on the web server follows the following pseudo-code:
\begin{algorithm}[H]
        \caption{The PIP Method in the client-server architecture}
        \label{algo:server}
        \begin{algorithmic}[c]
                \STATE At the Client side:
                \STATE\hspace{\algorithmicindent} Input parameters from HTML form
                \STATE\hspace{\algorithmicindent} HTTP request to compiled Fortran-90 code (Server side)
                \STATE At the Server side: 
                \STATE\hspace{\algorithmicindent} Compute physical map dimensions
                \STATE\hspace{\algorithmicindent} Generate random stars coordinates 
                \STATE\hspace{\algorithmicindent} Generate mass of stars
                \STATE\hspace{\algorithmicindent} Compute Dirichlet conditions 
                \STATE\hspace{\algorithmicindent} Compute common polygon areas
                \STATE\hspace{\algorithmicindent} Return 2D array with magnification map to HTML
                \STATE At the Client side (again):
                \STATE\hspace{\algorithmicindent} Javascript convolution in HTML (optional)
                \STATE\hspace{\algorithmicindent} Compute magnification probability ditributions
                \STATE\hspace{\algorithmicindent} Plot probability distributions and magnification map
        \end{algorithmic}
\end{algorithm}
Whereas most of the microlensing computations are made inside a Fortran-90 code on the server side, including the solution of the two-dimensional Poisson equation and the application of the inverse polygon approach, the convolution with a Gaussian source profile, the mean magnification, and the magnification probability distributions are calculated inside the pure HTML code. In order to run the microlensing generator, we installed it on an Apache Debian server, on an Intel Xeon$^{TM}$ E5-2620 CPU @ 2.10~Ghz with two cores. The web interface\footnote{\url{https://microlensing.overfitting.es}} allows the user to modify all the microlensing parameters and the physical size of the magnification map. The user can select the values for the convergence, shear, and smooth matter fraction using slide bars. The mass distribution of the microlenses in the lensing galaxy can be selected from three choices: constant (i.e. all lenses have equal mass), the Kroupa power-law distribution (with $\alpha=-1.3$), and the Salpeter power-law  distribution (with $\alpha=-2.35$). Finally, the physical size of the magnification map can cover 5, 10, 20, 50, or 100 Einstein radii with a length side of 500, 1000, or 2000 pixels. We invite any interested user to contact us for other sizes of the magnification patterns.

In principle, the magnification map will be generated for a point-like source (one pixel size), but it is possible to convolve the magnification map with a Gaussian source profile using the slide bar `source radius' in Einstein radius units (source size $\le 1$ $R_E$). The convolution is done inside the HTML code. As such, the convolution is a very fast procedure.

The magnification map is generated by pressing the button `Generate Map'. The mean magnification of the map will be shown in the corresponding box and compared with its theoretical expected value. The distribution of probability of the magnification map, normalised by its theoretical value, will appear in the block `Magnification Probability Distribution', and the magnification map itself will be in the block `Magnification Map', which  has to be selected. The generation of the magnification map will only take a few seconds. In fact, the time used on transferring the map to the browser will be longer than the computing time itself. The computing time is of the order of $10$ seconds.

Finally, the user has the possibility to download the generated magnification map to his/her local computer for subsequent analysis. By pressing the button `Save Map', a binary file is downloaded.  

\section{Conclusions}\label{sec:final}
The magnification maps are key tools for simulating quasar microlensing scenarios and obtaining  deep insights from quasar structure and lens mass distributions (e.g. \citealp{2020ApJ..905..7}). This knowledge helps in understanding the universe at a large scale. Additionally, microlensing effects in gravitationally lensed supernovae (SNe) have recently received quite a bit of attention (e.g. \citealp{2020A&A...633...162} and references therein), and these SN studies also require magnification maps. Realistic simulations imply the use of a large number of maps to include the proper motion of microlenses inside the lensing galaxy and, also, to cover a large area with a high definition in order to obtain good statistics on the behaviour of the source when moving through magnification patterns. New surveys will produce, in the very near future, many lens system candidates, and magnifications maps will play an important role in detailed lensing studies. For all these reasons, a public tool to produce a large amount of magnification patterns for quasar lensing (or SN) studies at the cost of a few seconds per map on a low-cost computer could prove to be a very useful tool for researchers. In this paper we provide such a tool.

The PIP method developed in this paper uses two different ideas: first, the two-dimensional Poisson solver for the deflection potential to reduce the method dependence on the number of microlenses and, second, the IPM method to reduce the dependence on the number of rays used in traditional ray-shooting approaches. The result of this combination is a method that produces magnification maps with the same accuracy as the ones produced with the other methods available in the literature, but in a much faster manner and without demanding computer resources.

We conclude with one final remark on the software presented here. The very fast speed in the production of magnification maps with our new method allows it to be run on a publicly accessible online server. But, obviously, the method is also ready for the production of a massive number of magnification maps either to explore a particular system, studying the movement of microlenses, or to build a large database to explore the large parameter space, including all the existing gravitational lens systems and those to come from forthcoming astronomical surveys (see also the GERLUMPH project\footnote{The GERLUMPH database at https://gerlumph.swin.edu.au (e.g. \citealp{2013MNRAS.434..832V}) contains magnification maps for many combinations of $\kappa$ and $\gamma$.}). We will explore both possibilities and invite interested researchers to contact the authors regarding possible collaborations.

\section*{Acknowledgements}{}
We thank Evencio Mediavilla for providing us with the IPM code for comparison purposes. We thank to the company ``Datacom Soluciones Internet Burgos S.L.'' (Burgos, Spain) for helping us in making the web server secure and running. We also thank an anonymous referee for his/her comments and questions that improved this paper. This research has been supported by the MINECO/AEI/FEDER-UE grant AYA2017-89815-P and University of Cantabria funds. RGM was also supported by TAILOR Grant \#952215, H2020-ICT-2019-3 and DataPol UMA-CEIATECH-07 funds at the University of M\'alaga.

\begin{appendix}
        \section{Overlap between the transformation onto the source plane of an image-plane unit cell and a source-plane pixel partially covered by it}\label{ap:intersect}
        
        The intersection of a parallelogram with a square is a polygon whose vertices are determined by the vertices of the parallelogram, the vertices of the square, and the intersection points between the sides of both geometrical objects. The intersection polygon is convex, and its area can be easily calculated when all vertices are ordered anticlockwise (see Sect.~\ref{sec:ipm}).
        
\begin{figure}[ht]
                \begin{center}
                        \includegraphics[width=\columnwidth]{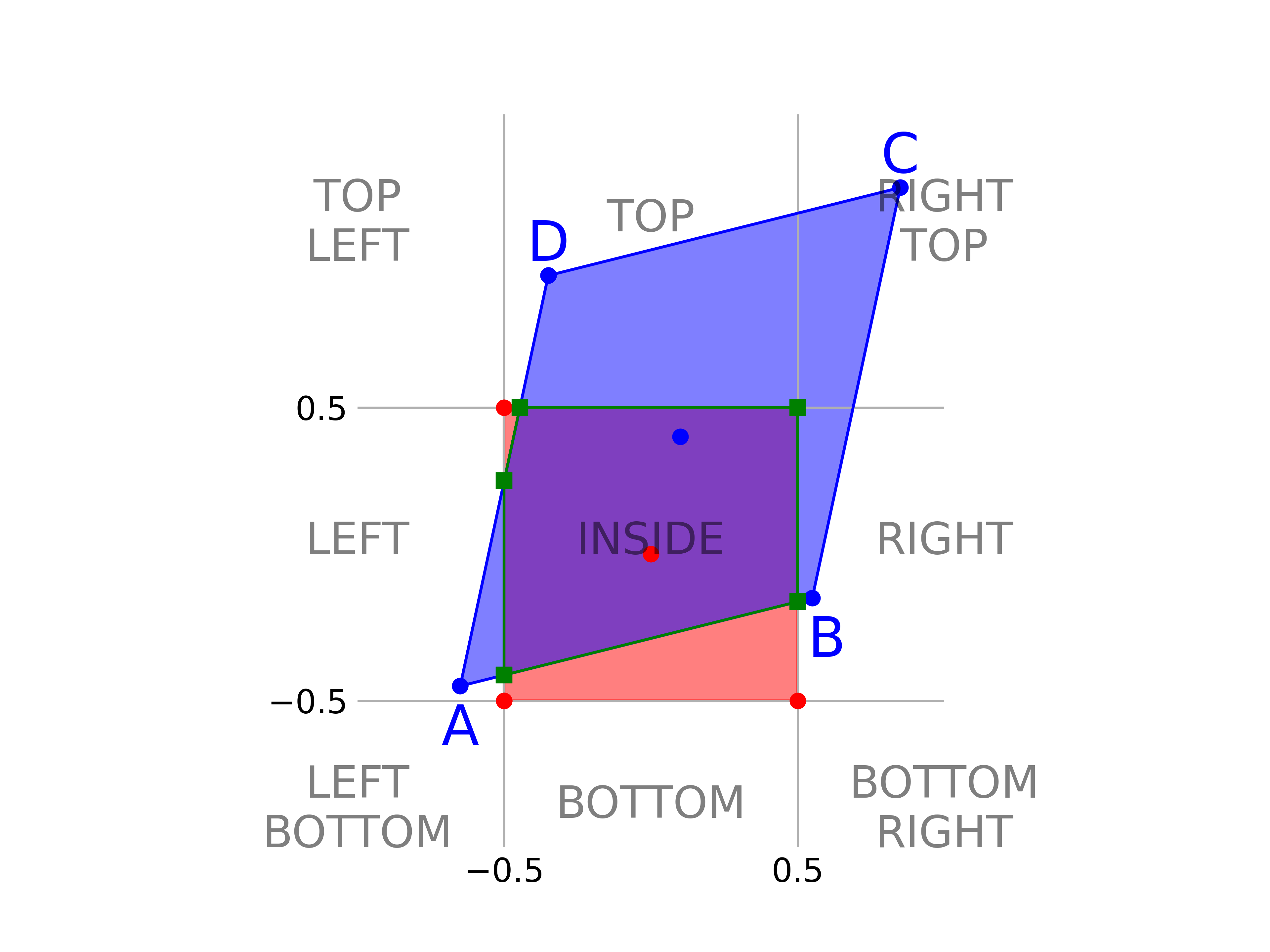}
                \end{center}
                \caption{Intersection between a transformed parallelogram and a unit squared cell in the source plane.}
                \label{fig:intersect}
        \end{figure}

        The key task is to find the intersection polygon, and there are some algorithms to do that. We followed the Maillot (1992) procedure, which is based on the Cohen-Sutherland line clipping algorithm. While Maillot (1992) considered the intersection of an arbitrary convex polygon with a rectangle, we focused on the particular case of a parallelogram (transformation onto the source plane of an image-plane unit cell) that partially covers a square (source-plane pixel). This allowed us to simplify the original algorithm and speed up calculations. A typical configuration is depicted in Fig.~\ref{fig:intersect}.     

        In Fig.~\ref{fig:intersect} the parallelogram (blue) has four vertices in anticlockwise order (A, B, C, and D), and each of them lies inside one of the nine labelled zones. One of these zones corresponds to the square pixel (red), and the other eight surround it. Additionally, the four sides of the parallelogram are characterised by two slopes:
        \begin{equation}
                s_1 = (y_2^\textnormal{B}-y_2^\textnormal{A}) / (y_1^\textnormal{B}-y_1^\textnormal{A})
        \end{equation}
        and
        \begin{equation}
                s_2 = (y_2^\textnormal{D}-y_2^\textnormal{A}) / (y_1^\textnormal{D}-y_1^\textnormal{A}).
        \end{equation} 
        
        To discuss the overlap between the parallelogram and the pixel centred at the origin of the coordinate system, we considered each side of the parallelogram in turn.

First, vertices A and B are located inside the LEFT and RIGHT zones, respectively. Thus, the AB side crosses both vertical sides of the pixel at points:                
        \begin{eqnarray*}
                \boldsymbol{y}^1 & = & (-0.5, ~y_2^\textnormal{A} + s_1 (-0.5 - y_1^\textnormal{A})) \\
                \boldsymbol{y}^2 & = & (+0.5, ~y_2^\textnormal{A} + s_1 (0.5 - y_1^\textnormal{A})).
        \end{eqnarray*}

Second, vertices B and C are within RIGHT and RIGHT TOP zones. The BC side does not intersect any pixel side, but the right-top vertex of the pixel is within the parallelogram and must be considered as a third reference point:
        \begin{eqnarray*}
                \boldsymbol{y}^3 & = & (+0.5, ~+0.5).
        \end{eqnarray*}
        
Third, vertices C and D are located in the RIGHT TOP and TOP zones. Hence, the CD side does not add any additional reference point.
        
        Fourth, vertex D is within the TOP zone, whereas vertex A is located inside the LEFT zone. In our particular case, the DA side crosses a horizontal and then a vertical side of the pixel at points:
        \begin{eqnarray*}
                \boldsymbol{y}^4 & = & (y_1^\textnormal{D} - (y_2^\textnormal{D} - 0.5)/s_2, ~+0.5) \\
                \boldsymbol{y}^5 & = & (-0.5, ~y_2^\textnormal{D} - s_2 (y_1^\textnormal{D} + 0.5)).
        \end{eqnarray*} 
         
        In the case we show in Fig.~\ref{fig:intersect}, the intersection polygon is defined by five final vertices at $\boldsymbol{y}^i, i = 1, ..., 5$ (small green squares).

        In a general case, there are 81 possible combinations between two parallelogram vertices lying inside one or two of the nine separate zones. For example, if both vertices are in the same zone, they do not contribute to the list of reference points unless they are included in the INSIDE region. For many other combinations, the algorithm either does not add new points to the reference list (e.g. LEFT BOTTOM and LEFT) or simply stops because the anticlockwise-ordered parallelogram and the pixel do not overlap (e.g. LEFT and TOP LEFT).   
        
\end{appendix}  
        

\begin{thebibliography}{}
\bibitem[Bartelmann(2003)]{2003astro.ph..4162B} Bartelmann, M., 2003, arXiv:astro-ph/0304162
\bibitem[Bate et al.(2010)]{2010NewAstr..15..725} Bate, N.~F., Fluke, C.~J., Barsdell, B.~R., Garsden, H., Lewis, G.~F., 2010, New Astronomy 15, 726
\bibitem[Chang and Refsdal(1979)]{1979Natur..282..561} Chang, K., Refsdal, S., Nature 282, 561
\bibitem[Chen et al.(2017)]{2017Astr.Comp..19..60} Chen, B., Kantowski, R., Dai, X., Baron, E., Van der Mark, P., 2017, Astronomy and Computing 19, 60
\bibitem[Cornachione et al.(2020)]{2020ApJ..905..7} Cornachione, M.~A., Morgan, C.~W., Burger, H.~R., et al., 2020, \apj, 905, 7
\bibitem[Corrigan et al.(1991)]{1991AJ....102...34C} Corrigan, R.~T., Irwin, M.~J., Arnaud, J., et al., 1991, \aj, 102, 34
\bibitem[Foltz et al.(1992)]{1992ApJ...386L..43F} Foltz, C.~B., Hewett, P.~C., Webster, R.~L., et al., 1992, \apjl, 386, L43
\bibitem[Gil-Merino et al.(1998)]{1998Ap&SS.263...47G} Gil-Merino, R., Goicoechea, L.~J., Serra, M., et al., 1998, \apss, 263, 47
\bibitem[Hockney \& Eastwood(1988)]{1988csup.book.....H} Hockney, R.~W., \& Eastwood, J.~W., 1988, Computer Simulations using Particles (Bristol: Hilger)
\bibitem[Irwin et al.(1989)]{1989AJ.....98.1989I} Irwin, M.~J., Webster, R.~L., Hewett, P.~C., et al., 1989, \aj, 98, 1989
\bibitem[Kayser et al.(1986)]{1986A&A...166...36K} Kayser, R., Refsdal, S., \& Stabell, R., 1986, \aap, 166, 36
\bibitem[Kochanek(2004)]{2004ApJ...605...58K} Kochanek, C.~S., 2004, \apj, 605, 58
\bibitem[Liebes(1964)]{1964PhRv..133..835L} Liebes, S., 1964, Physical Review, 133, 835
\bibitem[Mediavilla et al.(2006)]{2006ApJ...653..942M} Mediavilla, E., Mu{\~n}oz, J.~A., L\'opez, P., et al., 2006, \apj, 653, 942 
\bibitem[Mediavilla et al.(2011)]{2011ApJ...741...42M} Mediavilla, E., Mediavilla, T., Mu{\~n}oz, J.~A., et al., 2011, \apj, 741, 42
\bibitem[Paczynski(1986)]{1986ApJ...301..503P} Paczynski, B., 1986, \apj, 301, 503
\bibitem[Maillot(1992)]{1992ACM...11.3..276} Maillot P.-G., 1992, ACM Trans. Graph., 11, 3, 276
\bibitem[Popovic et al.(2006)]{2006AN....327..981P} Popovic, L.~C., Jovanovic, P., Petrovic, T. \& Shalyapin, V.~N., 2006, Astronomische Nachrichten, 327, 981
\bibitem[Schmidt \& Wambsganss(1998)]{1998A&A...335..379S} Schmidt, R. \& Wambsganss, J., 1998, \aap, 335, 379
\bibitem[Shalyapin(1995)]{1995ARep...39..594S} Shalyapin, V.~N., 1995, Astronomy Reports, 39, 594
\bibitem[Schneider et al.(1992)]{1992grle.book.....S} Schneider, P., 
Ehlers, J. \& Falco, E.~E., 1992, Gravitational Lenses (Berlin: Springer)  
\bibitem[Schneider \& Weiss(1987)]{1987A&A...171...49S} Schneider, P. \& Weiss, A., 1987, \aap, 171, 49
\bibitem[Suyu et al.(2020)]{2020A&A...633...162} Suyu, S.~H., Huber, S., Ca\~nameras, R. et al. 2020, \aap, 644, 162
\bibitem[Vernardos \& Fluke(2013)]{2013MNRAS.434..832V} Vernardos, G. \& Fluke, C.~J., 2013, \mnras, 434, 832 
\bibitem[Wambsganss(1990)]{1990PhDT.......180W} Wambsganss, J.\ 1990, Ph.D.~Thesis, Ludwig-Maximilians-Univ., Munich 
\bibitem[Wambsganss et al.(1990)]{1990ApJ...352..407W} Wambsganss, J., Paczynski, B., \& Katz, N., 1990, \apj, 352, 407
\bibitem[Webster et al.(1991)]{1991AJ....102.1939W} Webster, R.~L., Ferguson, A.~M.~N., Corrigan, R.~T., et al., 1991, \aj, 102, 1939
\bibitem[Witt \& Mao(1994)]{1994ApJ...429...66W} Witt, H.~J. \& Mao, S., 1994, \apj, 429, 66
\bibitem[Yonehara(1999)]{1999ApJ...519L..31Y} Yonehara, A., 1999, \apjl, 519, L31
\end{thebibliography}
\end{document}